\documentclass[journal]{IEEEtran}

\usepackage{caption} 
\usepackage{amsfonts}

\ifCLASSINFOpdf
   \usepackage[pdftex]{graphicx}
   \graphicspath{{../pdf/}{../jpeg/}}
   \DeclareGraphicsExtensions{.pdf,.jpeg,.png}
\else
\fi
\usepackage{amsmath}

\usepackage{multirow}
\usepackage{algorithmic}

\usepackage{array}
\begin{document}
%
\title{Image Classification using Sequence of Pixels}
%
%
%

\author{Gajraj Kuldeep
       
\thanks{ Gajraj Kuldeep is with the Department of Engineering, Aarhus University,
	8000 Aarhus C, Denmark. Email:  \textbf{gkuldeep@eng.au.dk}}
}

%
%

\markboth{}%
{Shell \MakeLowercase{\textit{et al.}}: Sequencial Image Classification}
%



\maketitle

\begin{abstract}
This study compares sequential image classification methods based on recurrent neural networks. We describe methods based on recurrent neural networks such as Long-Short-Term memory(LSTM), bidirectional Long-Short-Term memory(BiLSTM) architectures, etc. We also review the state-of-the-art sequential image classification architectures. We mainly focus on  LSTM, BiLSTM, temporal convolution network, and independent recurrent neural network architecture in the study. It is known that RNN lacks in learning long-term dependencies in the input sequence. We use a simple feature construction method using orthogonal Ramanujan periodic transform on the input sequence. Experiments demonstrate that if these features are given to LSTM or BiLSTM networks, the performance increases drastically. 

Our focus in this study is to increase the training accuracy simultaneously reducing the training time for the LSTM and BiLSTM architecture, but not on pushing the state-of-the-art results, so we use simple LSTM/BiLSTM architecture. We compare sequential input with the constructed feature as input to single layer LSTM and BiLSTM network for MNIST and CIFAR datasets. We observe that sequential input to the LSTM network with 128 hidden unit training for five epochs results in training accuracy of 33\% whereas constructed features as input to the same LSTM network results in training accuracy of 90\% with $\frac{1}{3}$ lesser time. 
\end{abstract}

\begin{IEEEkeywords}
recurrent neural network(RNN), deep independent RNN, Long-Short-Term memory, Bidirectional LSTM, orthogonal Ramanujan periodic transform.
\end{IEEEkeywords}

\IEEEpeerreviewmaketitle

\section{Introduction}

\IEEEPARstart{I}{mage} classification is a fundamental problem in computer vision and machine learning, it is used to classify images into predefined class of objects. In sequential image classification tasks, images are processed as long sequences, one pixel at a time. It is different from other image classification problems because there complete image is available for processing.  Deep learning techniques have been well developed and extensively used to classify images: there are several types of architecture for deep learning, such as recurrent neural network (RNN), convolution neural network (CNN) and deep neural network (DNN).  Development of recurrent neural network is contributed to the authors \cite{rnn1,rnn2} and \cite{rnn3},  It is widely believed that  for sequential data RNNs perform better than CNN and DNN. 

In this study, first we review the state-of-the-art network architecture for sequential image classification and in the second part we introduce a new method to construct features with the aim of reducing training time and increasing testing and training accuracy. 

The paper is organized as follows: Section II contains the basics of recurrent neural network and also describe deep independent RNN and temporal convolution network. Section III describes feature construction method Section IV. contains performance results for state-of-the-art networks and also contains results for LSTM and BiLSTM architectures. Finally, Section V concludes the paper. 

\textit{Notations:} In this paper, all boldface uppercase letters such as $\mathbf{X}$ represent matrices. All boldface lowercase  letters such as, $\mathbf{x}$   represent vectors. $\mathbf{x}^T$ is transpose of $\mathbf{x}$. $x^t$ is a time sample at $t$ of sequence.$\mathbf{x}^t$ represents vector values at time $t$.

\section{Recurrent Neural Network }
\begin{figure}[htbp]
	
		\includegraphics [width=7.0cm]{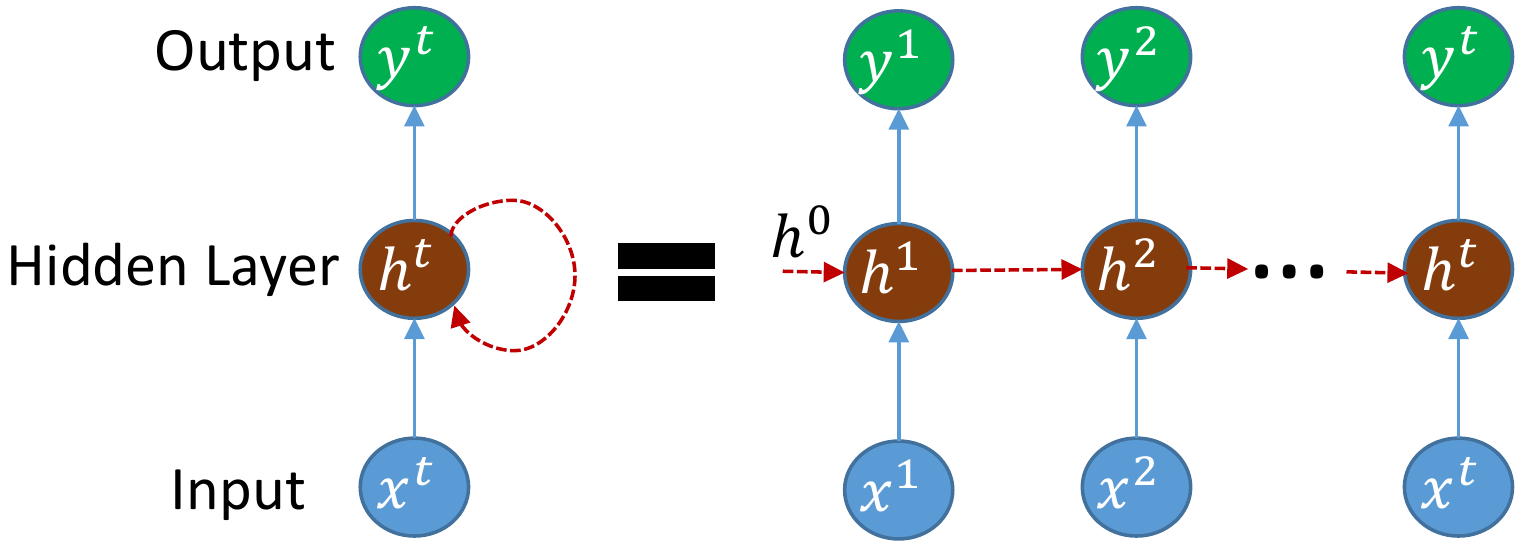}
		\caption{A single node recurrent neural network and its time evolution.}
		\label{rnn}
\end{figure}

RNNs are used to find patterns in time series data, such as stock market prediction, language modeling, text and speech generation etc. It learns the dependencies in a time series data through the feedback connections. But it suffers from the gradient vanishing and exploding problem. A simple RNN with single node is shown in Fig. \ref{rnn}. The output of the hidden layer is stored in the memory layer, which will be used for future modeling of the input.  
The time used in RNN is discretized; hence the feedback is delayed by the corresponding time unit. From Fig. \ref{rnn}, the mathematical expression for the update of hidden state is given as,

\begin{equation}
\mathbf{h}^t=\sigma(\mathbf{W}_h\mathbf{h}^{t-1}+\mathbf{W}_x\mathbf{x}^{t}+\mathbf{b}), \label{ernn}
\end{equation}
where $\mathbf{x}^t\in\mathbb{R}^N$ and $\mathbf{h}^t\in\mathbb{R}^M$ are input and hidden state vectors at time $t$. Similarly, weight matrices for input and recurrent input are $\mathbf{W}_x\in\mathbb{R}^{M\times N}$ and $\mathbf{W}_h\in\mathbb{R}^{M\times M}$, respectively. It should be noted that $\sigma$ is an element wise activation function. $\mathbf{b}\in\mathbb{R}^M$ is a bias vector. The output of the hidden layer at time $t$ is given as,
\begin{equation}
\mathbf{y}^t=\mathbf{h}^t.
\end{equation}

For long data sequences the gradients become smaller with each iteration which creates problem in the learning. This problem is called gradient vanishing problem in the literature. RNNs also have problem of gradient explode. The problem of gradient explode occurs when the gradient values become very high. Different RNN architectures were proposed to avoid these problems and also to learn long term dependencies in the data. Concept of Long Short-Term Memory (LSTM) was introduced by Hochreiter and Schmidhuber \cite{lstm1,lstm2} to avoid gradient exploding and vanishing problems. They replaced the traditional nodes with the memory cell in the hidden layer. The introduction of these memory cells facilitated in learning of long term dependencies in the data. The LSTM network uses only past input values to determine output at any point in the data. Since the LSTM network is dependent on the past inputs, a new network architecture was designed to cater past and future values of input to determine the output at any point in the data.This architecture is called Bidirectional Recurrent Neural Network (BRNN) \cite{brnn1}. It is argued that LSTM networks suffer from gradient decay over layers; hence construction and training of deep LSTM network is practically difficult \cite{Indrnn1,Indrnn2}.
\subsection{Deep independent RNN}
\begin{figure}[htbp]
	\centering
	\includegraphics [width=8.0cm]{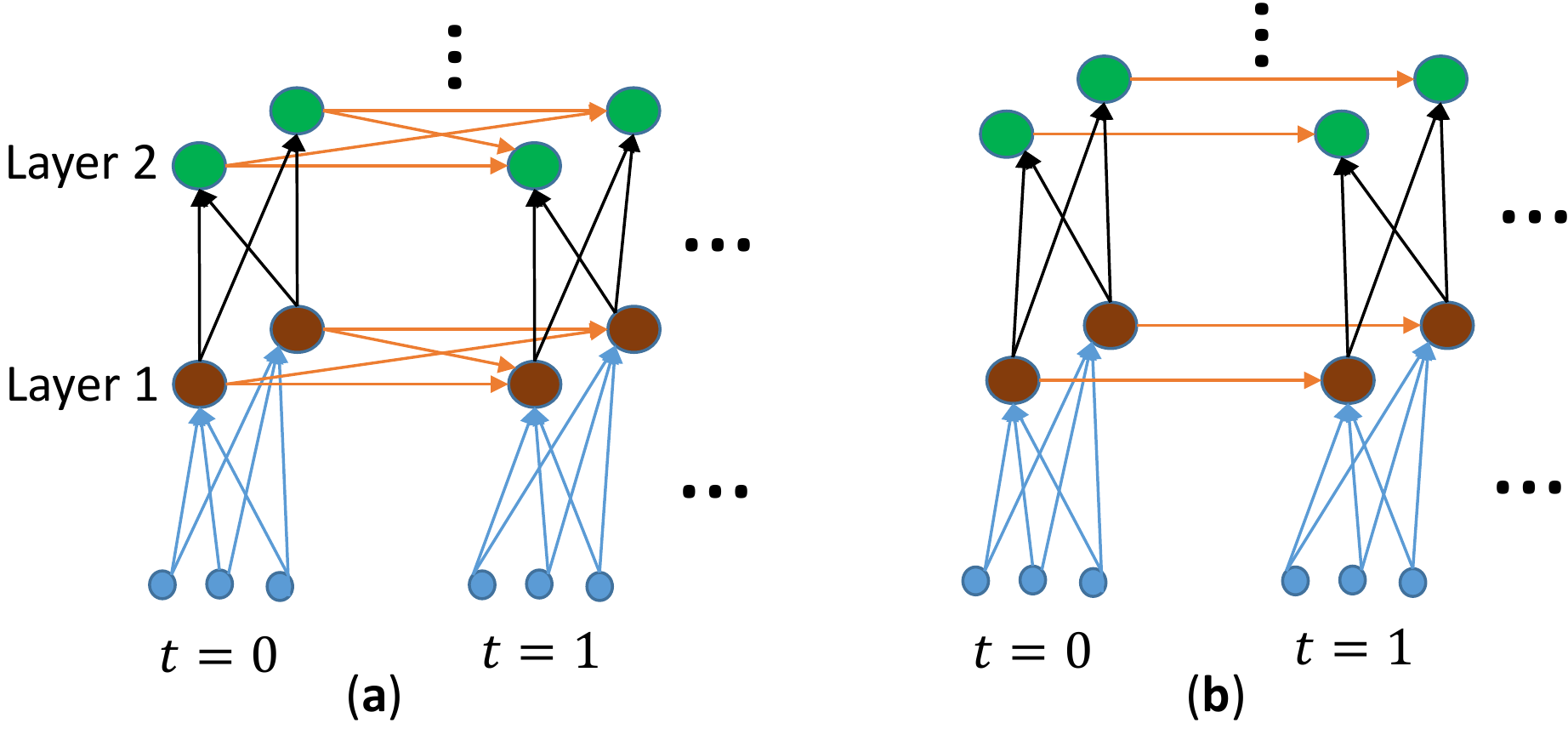}
	\caption{Illustration of multi-layered RNN and IndRNN. (a) The multi-layered RNN and its evolution along time. (b) IndRNN and its evolution along time.}
	\label{indrnn1}
\end{figure}

A new architecture \cite{Indrnn1,Indrnn2} was proposed based on Eq. \ref{ernn} and named as deep independent recurrent neural network (IndRNN) (\textit{Source code: https://github.com/Sunnydreamrain/IndRNN\_pytorch}). In this architecture recurrent connection were made to behave independently by using diagonal recurrent weight matrix.
 
\begin{equation}
\mathbf{h}^t=\sigma(\mathbf{w}_h\odot\mathbf{h}^{t-1}+\mathbf{W}_x\mathbf{x}^{t}+\mathbf{b}), \label{eindrnn}
\end{equation}
where $\mathbf{w}_h$ is a column vector.  IndRNN is a combination of  evolution in time direction as well as in layers as can be seen in Fig. \ref{indrnn1}. 
It can be observed from Fig. \ref{indrnn1} that only difference between the multi-layered RNN and IndRNN is that the recurrent connections, in the multi-layered RNN recurrent connections are shared whereas in IndRNN they are independent.  
\subsection{Temporal convolution network}
Sequential image classification can also be performed using temporal convolution network (TCN) architecture.  Sequence modeling using TCN is described in \cite{rtcn}, here we present the basic elements of TCN (\textit{Source code: https://github.com/locuslab/TCN}). The basic element of TCN is full convolution network (FCN). In TCN one-dimensional FCNs are applied in the causal setting, it means that that there is no information leakage from future to the past. It uses dilated convolutions to capture long term dependencies. TCN looks similar to the RNN in terms of sequential analysis. However, there is one difference between RNN and TCN in terms of storage of inputs, in the RNN only present input sample is required other past input samples are not required where as in the TCN current input sample as well as past all input samples are required.  

Sequence modeling using TCN can be formalized in the following way. Let $x^0,x^1\dots,x^T$ be an input sequence and corresponding output sequence is $y^0,y^1\dots,y^T$. The only constraint to produce output $y^t$ at a time $t$ is that the network can use only input sequence $x^0,x^1\dots,x^t$ up to time $t$. To accomplish this modeling  causal convolutions are applied on the input. Simple causal convolution is only able to look back at a history with size linear in the depth of network, which requires large number of layers. This is challenging where long dependency in data is required, which brings us to the concept of dilated convolutions\cite{dcon1,dcon2}.  

Mathematically the dilation convolution operation on a sequence $\mathbf{x}$ for operation $R$ on the $k^{th}$ time using filter $\mathbf{f}=\{f^0,f^1,\dots,f^{L-1} \}$ is given as,
\begin{equation}
R^k=\mathbf{x}*_df^{(k)}=\sum_{i=0}^{L-1}f^ix^{k-di}, \label{edc}
\end{equation} 
where $d$ is dilation and $L$ is the filter length. For $d=1$ Eq. \ref{edc} is a normal convolution. As the $d$ increases the receptive field also increases. Receptive field is a CNN concept and it is defined as the region in the input space which is affected by the CNN operation. Therefore, here it means regions affected by the dilation convolution. From Eq. \ref{edc} we can observe that the receptive field can be increased by increasing either $L$ or $d$.  Authors \cite{rtcn} increases the $d=O(2^i)$ exponential at the $i^{th}$ layer of the network. By this way, it can be ensured that every input sample is hit by some filter which ensures large learning depth.

The concept of deep residual learning was explored to mitigate the problem of increase in training and testing errors by increasing the number of hidden layers in deep neural network \cite{rtcnrc}.  It has been shown that increasing layer using residual network can improve performance.
%
%
Since the number of layers in TCN is high, residual network architecture is adopted for better performance. Residual block has a identity transformation for the input and many series of transformation on the input and the output is represented as,
\begin{equation}
\mathbf{o}=\sigma(\mathbf{x}+F(\mathbf{x})),
\end{equation}
where $\sigma$ is a activation function and $F$ is a composition of series transformations. Network depth, filter size, and $d$ determine the effective receptive field. Therefore, it is imperative to use residual network for stabilization in the TCN networks.  It has two dilation convolution layers followed by weightnorm layers \cite{weight}, rectified linear unit \cite{ReLU}, and dropout layer \cite{dropout}.
In the ResNet input is applied with identity transform whereas in the TCN when the output and input have the different width then $1\times1$ convolution is used to match the shape.

\section{Time parallelism in RNN}
\begin{figure}[htbp]
	\centering
	\includegraphics [width=8.0cm]{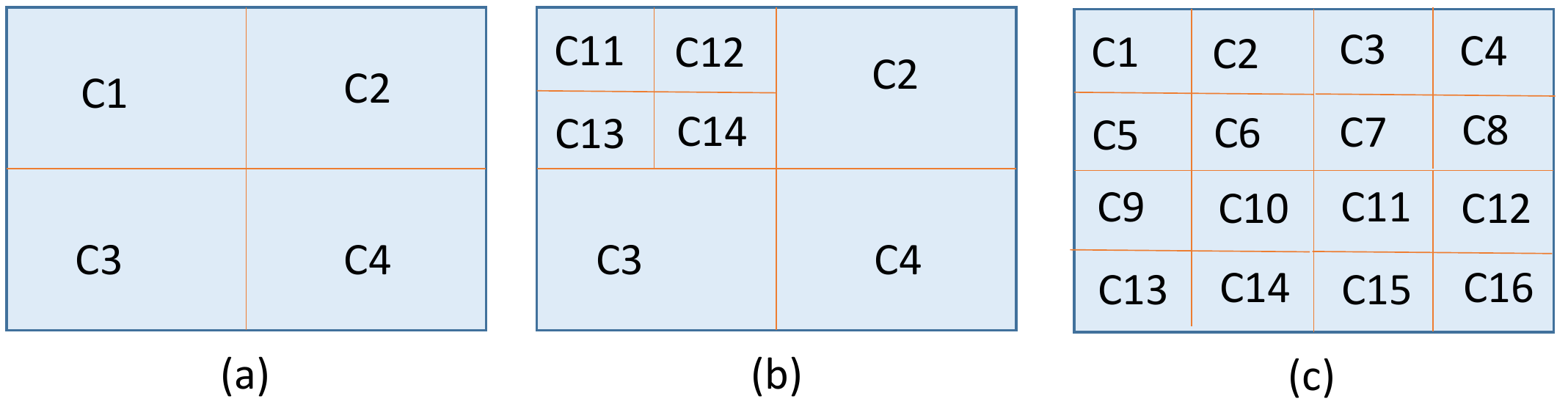}
	\caption{Illustration of Wavelet transform to generate multiple features. (a) Application of Haar wavelet to get 4 features  . (b) Application of Haar wavelet to get 7 features.(c) Application of ORPT to get 16 features. }
	\label{fwavelet}
\end{figure}

It is common practice to extract features from the input then apply the RNN on the extracted features. As in the paper \cite{cnnrnn}, CNN is used for feature extraction and RNN is used for  classification. One can use CNN for feature extraction but it comes with high computational cost. Here we propose simple way to construct features for image data. RNNs are sequencial in nature means the network accepts input sample by sample. Therefor for image data the input sequence becomes very long assuming an image of size $256\times256$ the input sequence length becomes 65536. Which requires network to be very stable during such a long training session. To reduce the RNNs complexity, we introduce the time parallelism in the input. We know that wavelets \cite{wavelet} are used to analyze the images in different resolutions. For example if we apply Haar wavelet to an image we get four features as shown  in Fig. \ref{fwavelet}. 
If the Haar wavelet is applied on an image we get 4 features corresponding with equal sizes. For getting more number of features Haar wavelet can be applied again on the low frequency channel which results in 7 features if Haar wavelet is applied two times as shown in Fig. \ref{fwavelet}(b). However, this results into unequal sizes of the features which requires padding to make all features equal size. To avoid this, we use Ramanujan orthogonal periodic (ORPT) transform \cite{orpt1} to get equal size features as shown in Fig. \ref{fwavelet}(c). The ORPT is a integer based multichannel wavelet transform. 
  Application of ORPT-based analysis matrix on an image of size $N\times N$, where $N$ is divisible by integer $q$ will result in $q^2-1$ detail components and one average component in transform domain.
  ORPT is defined using Ramanujan sums \cite{ramanujan} which are family of trigonometric sum. Ramanujan sums are give as,
  \begin{equation}
  c_q(n) = \sum_{\begin{subarray}{c}
  	k=1\\
  	(k, q) =1\\
  	\end{subarray}}^{q}\exp^{\frac{j2\pi kn}{q}}
  \end{equation}
  where $(k,q)=1$ implies that k and q are relatively prime. It has been shown that these sequences behaves as discrete derivative \cite{RS}. It is also observed that well known derivative kernels such as Sobel, Prewitt, Laplacian etc. are part of these generalized derivatives. The ORPT transform is defined using sparse Ramanujan sequence, which is given below,
  Let us denote, for any prime $q$,
  \begin{equation}
  c^{k}_q(n) = u(((n))_q-k)c_q(n-k)-k\delta(((n))_q-k)
  \end{equation}
  where $((n))_q$ is $n$ mod $q$,\\\\ $ u(n-k) $ =$\left\{\begin{array}{cc}
  1 & n \ge k\\
  0 & else
  \end{array}\right.$ and $\delta(n-k)$ =$\left\{\begin{array}{cc}
  1 & n = k\\
  0 & else
  \end{array}\right.$\\ \\
  for $0 \le k < q-1$.

  Any arbitrary signal $x(n)$, of length $N$ transform representation is given as,
  \begin{eqnarray}
  x(n)  = \sum_{d_i | N} \sum_{j_1=0}^{(p_{i1}^{r_{i1}-1}-1)}  \hdots \sum_{j_m=0}^{(p_{im}^{r_{im}-1}-1)} \sum_{k_1=0}^{\phi(p_{i1})-1}  \hdots \sum_{k_m=0}^{\phi(p_{im})-1} \nonumber \\\beta_{d_i;j_1 \hdots ,j_m;k_1, \hdots ,k_m}  {c^{k_1}_{p_{i1}}(\frac{n}{p_{i1}^{r_{i1}-1}}-j_1)}   \hdots  {c^{k_m}_{p_{im}}(\frac{n}{p_{im}^{r_{im}-1}}-j_m)} \label{ORPT:}
  \end{eqnarray}
  
  $\beta_{d_i;j_1,j_2 \hdots ,j_m;k_1,k_2 \hdots ,k_m} $ are  ORPT coefficients of $x(n)$ which can be represented as : \\
  
  $<x(n),({c^{k_1}_{p_{i1}}(\frac{n}{p_{i1}^{r_{i1}-1}}-j_1)}   \hdots  {c^{k_m}_{p_{im}}(\frac{n}{p_{im}^{r_{im}-1}}-j_m)})>$\\\\ and  $d_i$'s are divisor of $N$ and each $d_i$ is of the form $p_{i1}^{r_{i1}}p_{i2}^{r_{i2}} \hdots p_{im}^{r_{im}}.$
  
  Eq. \ref{ORPT:} looks complicated but it generates integer matrix for any positive integer $N$. For $N=6$ the ORPT analysis matrix is given as,
  \begin{eqnarray*}
  	\mathbf{R}_{6} = \left[\begin{array}{rrrrrr}1 & 1 & 2 & 0 & 2 &0\\
  		1 & -1 & -1 & 1 & 1 &-1\\
  		1 & 1 & -1 & -1 & -1 &-1\\
  		1 & -1 & 2 & 0 & -2 &0\\
  		1 & 1 & -1 & 1 & -1 &1\\
  		1 &  -1 & -1 & -1 & 1 &1\end{array}\right]\\
  \end{eqnarray*}
  
  To apply ORPT on an image first a divisor $d$ is selected then the corresponding $\mathbf{R}_d$ matrix is constructed. After constructing $\mathbf{R}_d$ a wavelet analysis matrix is generated which can be represented as $\mathbf{B}$. To get $d^2$ features the wavelet analysis matrix is applied on the image in the following,
  \begin{equation}
  \mathbf{Y=BXB}^T,
  \end{equation}
  where $\mathbf{X}$ is an input image and $\mathbf{Y}$ is a transformed image. The features are called independent because the columns of the ORPT forms orthogonal basis.
  
\begin{figure}[htbp]
	\centering
	\includegraphics [width=8.0cm]{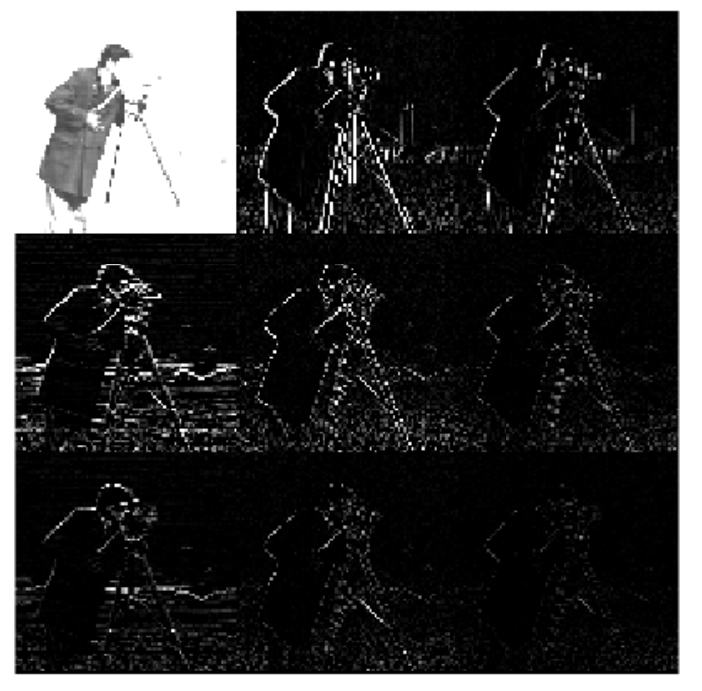}
	\caption{Illustration of generation of 9 equal size features using ORPT.  }
	\label{orpt3}
\end{figure}

   For example: cameraman image of size $255\times255$ the application of ORPT  for divisor 3 is shown in Fig. \ref{orpt3} to get 9 features of equal size. From the figure it can be observed that it has 8 detail (high frequency) features and one average (low frequency) channel.

\begin{figure}[htbp]
	\centering
	\includegraphics [width=9.0cm]{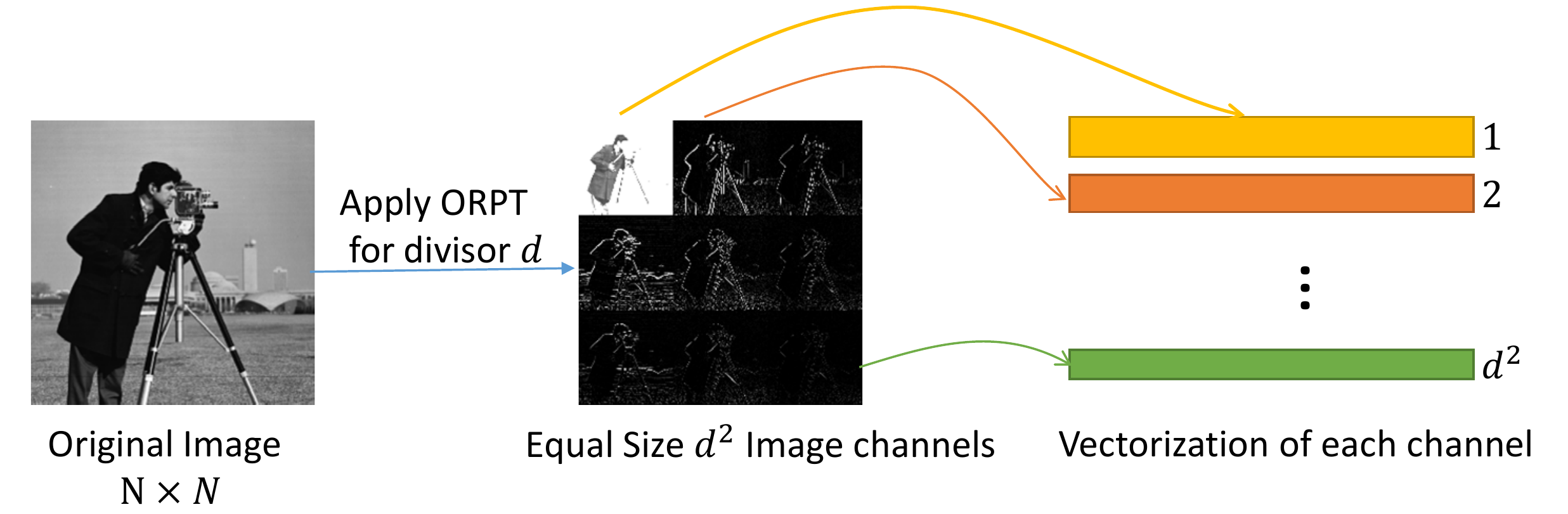}
	\caption{Independent channel construction using ORPT. }
	\label{feature}
\end{figure}

 The feature extraction using ORPT is given in Fig. \ref{feature}
To construct features, first divisor $d$ is selected then ORPT $\mathbf{R}_d$ is constructed and based on the constructed matrix $\mathbf{R}_d$ a wavelet analysis matrix is constructed, which results into multi-resolution of the image. The constructed multi-resolution image contains $d^2$ channels, $d^2$ features are constructed by vectorizing these channels.

\section{Performance Evaluation}
In this section, we compare the state-of-the-art algorithm for sequential image compression. Then we also compare the basic LSTM and BiLSTM architecture using time parallelism in the input for various divisors.
\subsection{Datasets} MNIST dataset \cite{MNIST} is widely used to test the classification performance of the neural network. MNIST contains images of digits from 0 to 9. Generally RNN is tested with sequential MNIST which is nothing but constructing a sequence particular image. We construct multiple sequences for testing the RNN under consideration. MNIST images are of sizes $28\times28$ there are 60000 training images and 10000 testing images.

We also conducted more analysis on these algorithms using CIFAR-10 dataset \cite{cifar}, which contains 50000 training images and 10000 testing images in 10 classes. Each image is a size of $32\times32$.

\subsection{Experimental setup } We performed our experiments on i7 CPU with 16GB RAM computer. We also used Google Colab GPU to verify a few results. All results on LSTM and BiLSTM are performed on computer using MATLAB.
\subsection{Performance evaluation of state-of-the-art networks}
Sequential image classification is very challenging because the sequence from the image is very long. Hence the network should be able to learn long term dependencies without the problem of gradient vanishing and exploding. In the literature, sequential image classification is demonstrated using two famous datasets, which are MNIST and CIFAR-10 as shown in the Table \ref{tab1}. For MNIST dataset sequentian image classification is comparable to CNN based classification methods. However for CIFAR-10 dataset the sequential image classification using RNN is not at par.
\begin{table}[htbp]
	\begin{tabular}{c|c|c|c|c|}
		\cline{2-5}
		& IndRNN\cite{Indrnn2} & TCN\cite{rtcn} & Transformer\cite{trans} & \begin{tabular}[c]{@{}c@{}}Trellis \\ Network\cite{tcn}\end{tabular} \\ \hline
		\multicolumn{1}{|c|}{\begin{tabular}[c]{@{}c@{}}Sequential \\ MNIST\end{tabular}} & 99.48 & 99 & 98.4 & 99.2 \\ \hline
		\multicolumn{1}{|c|}{CIFAR-10} & - & - & 62.3 & 73.4 \\ \hline
	\end{tabular}
	\caption{Comparison of test accuracy for MNIST and CIFAR-10 datasets for state-of-the-art networks for sequential image classification.}
	\label{tab1}
\end{table}

\subsection{Results for Time parallelism in RNN} 
\begin{figure}[htbp]
   	\centering
   	\includegraphics [width=9.0cm]{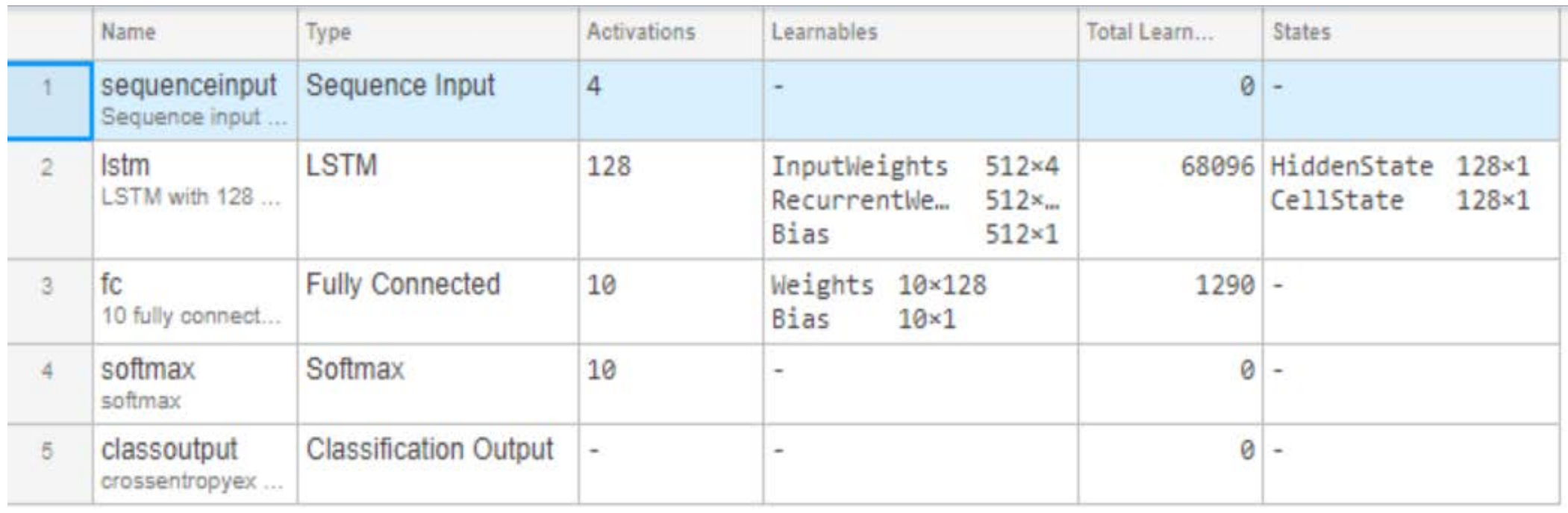}
   	\caption{Network analysis for features constructed using $d=2$ for single layer LSTM network.}
   	\label{rnet}
   \end{figure}

\begin{figure}[htbp]
	\centering
	\includegraphics [width=9.0cm]{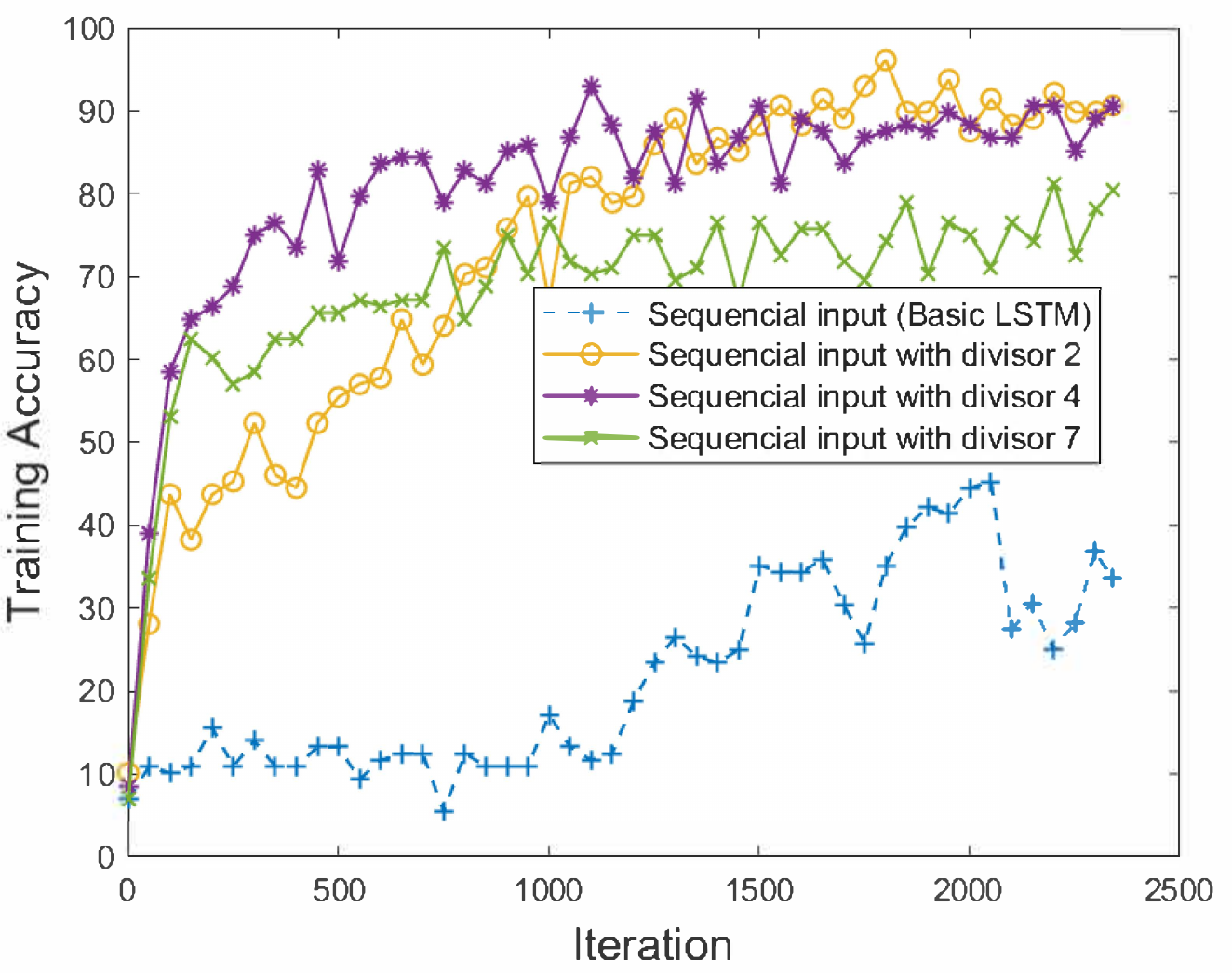}
	\caption{Comparison of training accuracy for various input divisors for single layer LSTM network on MNIST dataset.}
	\label{featureLTMS}
\end{figure}

\begin{figure}[htbp]
	\centering
	\includegraphics [width=9.0cm]{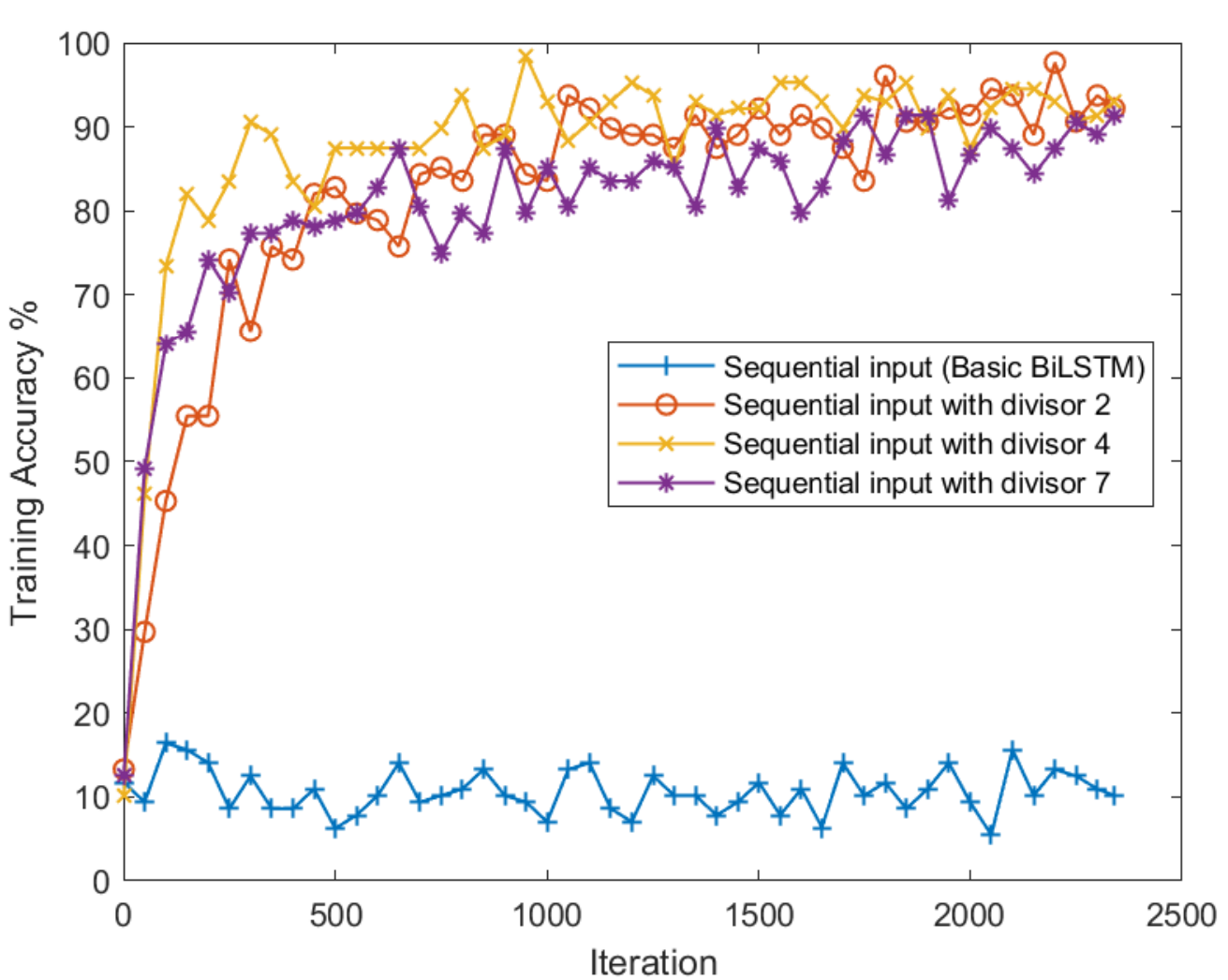}
	\caption{Comparison of training accuracy for various input divisors for single layer BiLSTM network on MNIST dataset.}
	\label{featureBiLSTM}
\end{figure}

\begin{figure}[htbp]
	\centering
	\includegraphics [width=9.0cm]{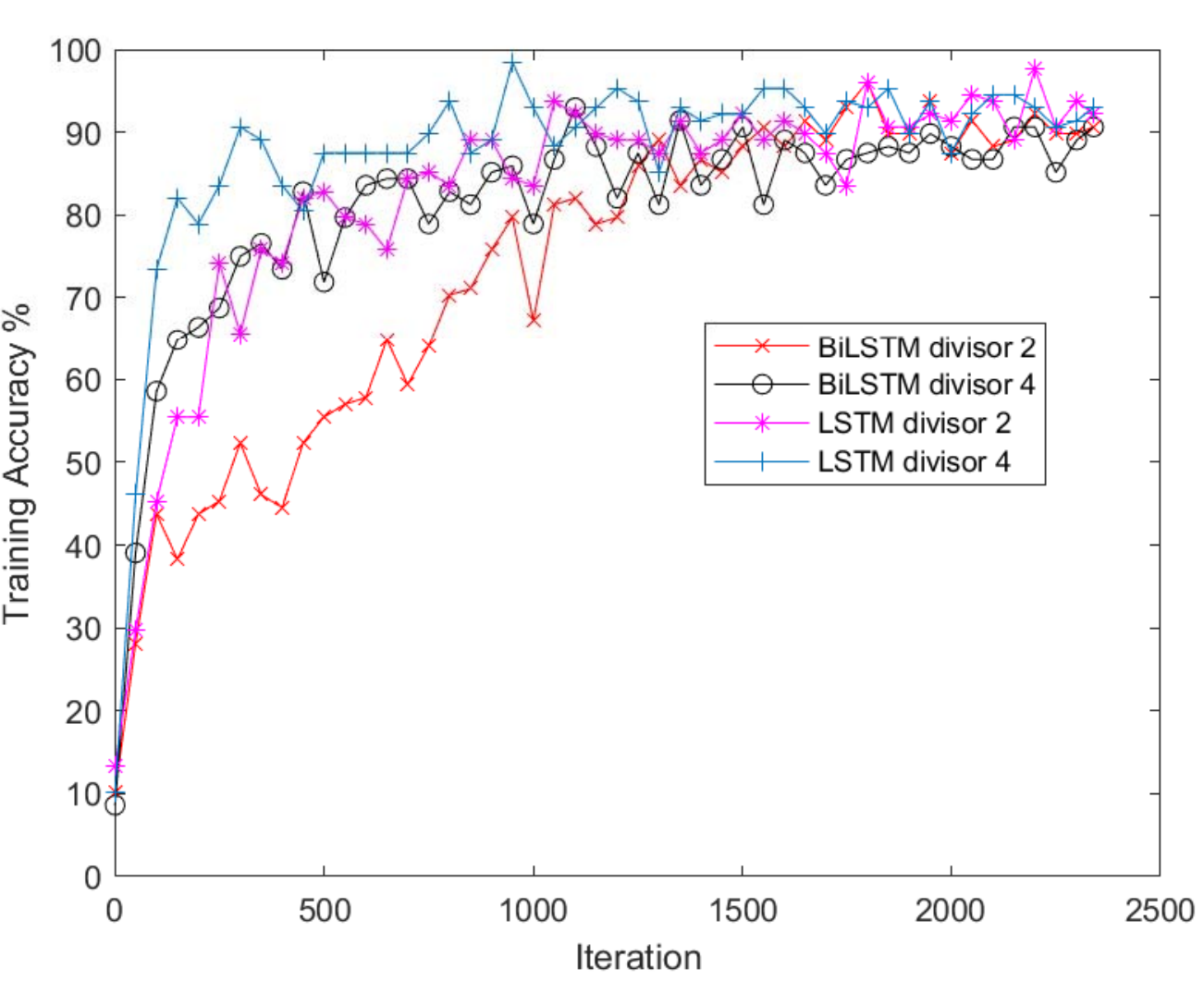}
	\caption{Comparison of training accuracy for single layer BiLSTM and LSTM network using divisors $2$ and $4$ on MNIST dataset.}
	\label{Comfeature}
\end{figure}

To verify the effect of using multiple features as sequential inputs to the network under consideration, we use single layer of LSTM or BiLSTM with 128 hidden states. MNIST images are of the size $28\times28$ and the divisors,$d$, of $28$ are $1,2,4,7,14$, and $28$. $d=1$ corresponds to general sequential input. As we know that the for divisor $d$ we get $d^2$ features. Network analysis of used network for $d=2$ is shown in Fig. \ref{rnet}.  
From the figure, it can also be observed that total trainable parameters for this network are 70K approximately. Training performance for this single layer LSTM network for different possible  divisors $d$   of $28$ are shown in Fig. \ref{featureLTMS}. 
It can be observed that the training performance increases with the $d$ as compared to the basic sequential input ($d=1$). However, we observe that the large $d$ the network training performance is not as good as for low divisors as this can also be observed from Fig. \ref{featureLTMS}. We have also analyzed signal layer BiLSTM network for various possible divisor as shown in Fig. \ref{featureBiLSTM}. For the BiLSTM network also the training performance increases with the introduction of features.    
We also compare single layer LSTM and BiLSTM network for $d=2$ and $d=4$ as shown in Fig. \ref{Comfeature}.
\begin{table}[htbp]
	\centering
	\begin{tabular}{l|l|l|l|l|}
		\cline{2-5}
		& \multicolumn{4}{l|}{LSTM} \\ \hline
		\multicolumn{1}{|l|}{Divisor} & 1 & 2 & 4 & 7 \\ \hline
		\multicolumn{1}{|l|}{\begin{tabular}[c]{@{}l@{}}Time in\\ minutes\end{tabular}} & 158 & 45 & 15 & 5 \\ \hline
		\multicolumn{1}{|l|}{\begin{tabular}[c]{@{}l@{}}Training \\ accuracy \%\end{tabular}} & 33.19 & 90.63 & 90.6 & 80.47 \\ \hline
		\multicolumn{1}{|l|}{\begin{tabular}[c]{@{}l@{}}Test \\ accuracy \%\end{tabular}} & 35.15 & 92.51 & 92.37 & 86.23 \\ \hline
	\end{tabular}
	\caption{Comparison of time taken, training accuracy, and test accuracy for various input divisors for single layer LSTM network for 2340 iterations on MNIST dataset.}
	\label{tab2}
\end{table}
Timing for the training is also compared for single layer LSTM and BiLSTM network, which are shown in the Table \ref{tab2} and \ref{tab3}. Form these tables, it can be observed that the training time reduces by $\frac{1}{3}$ with increase in testing and training accuracy. 
\begin{table}[]
	\centering
	\begin{tabular}{l|l|l|l|l|}
		\cline{2-5}
		& \multicolumn{4}{l|}{BiLSTM} \\ \hline
		\multicolumn{1}{|l|}{Divisor} & 1 & 2 & 4 & 7 \\ \hline
		\multicolumn{1}{|l|}{\begin{tabular}[c]{@{}l@{}}Time in\\ minutes\end{tabular}} & 447 & 68 & 21 & 7 \\ \hline
		\multicolumn{1}{|l|}{\begin{tabular}[c]{@{}l@{}}Training \\ accuracy \%\end{tabular}} & 10.16 & 92.19 & 92.97 & 91.41 \\ \hline
		\multicolumn{1}{|l|}{\begin{tabular}[c]{@{}l@{}}Test \\ accuracy \%\end{tabular}} & 12.13 & 93.51 & 92.25 & 90.43 \\ \hline
	\end{tabular}
	\caption{Comparison of time taken, training accuracy, and test accuracy for various input divisors for single layer BiLSTM network for 2340 iterations on MNIST dataset.}
	\label{tab3}
\end{table}

CIFAR-10 database contains color images of the size $32\times32$. The divisors of $32$ are $1,2,4,8,16$, and $32$. For a particular divisor $d$, the number of features are $3d^2$ for a color image. For the CIFAR-10 we observed similar performance improvement as in MNIST. 


\section{conclusion}
In this study, we reviewed the state-of-the-art deep learning architectures for sequential image classification. These network architectures were introduced either to avoid gradient exploding and vanishing problem or to enhance the long term learning capabilities. We took feature-based approach to enhance the long term learning of simple LSTM and BiLSTM network with reduced learning time as well as increasing the training and testing accuracy. We verified our approach using MNIST and CIFAR-10 datasets.

\end{document}